\documentclass[aip,
 amsmath,amssymb,
 reprint,
]{revtex4-1}
\usepackage{dcolumn}
\usepackage{bm}
\usepackage[utf8]{inputenc}
\usepackage[T1]{fontenc}
\usepackage{mathptmx}
\usepackage{graphicx}
\usepackage{amsmath,amssymb}
\usepackage{color}
\usepackage{textcomp}
\usepackage{float}

\begin{document}
\preprint{AIP/123-QED}
\title{\textcolor{black}{Computation of the chemical potential and solubility of amorphous solids}}
\author{H. A. Vinutha}
 \altaffiliation{Corresponding author}
\email{vh327@cam.ac.uk}
\affiliation{Institute of Physics, Chinese Academy of Sciences, Beijing, China}
\affiliation{Yusuf Hamied Department of Chemistry, University of Cambridge, Cambridge, UK}
\author{Daan Frenkel}
\email{df246@cam.ac.uk}
\affiliation{Yusuf Hamied Department of Chemistry, University of Cambridge, Cambridge, UK}
\date{\today}

\begin{abstract}
{\color{black}Using a recently developed  technique to estimate the equilibrium free energy of glassy materials, we explore if equilibrium simulation methods can be used to
estimate the solubility of amorphous solids.  
As an illustration, we compute the chemical potentials of the constituent particles of a two-component Kob-Andersen model glass former. 
To compute the chemical potential for different components, we combine the calculation of the overall free energy of the glass with a calculation of the  chemical potential difference of the two components.  
We find that the standard method to compute chemical potential differences by thermodynamic integration yields not only a wide scatter in the chemical potential values but, more seriously, the average of the thermodynamic integration results is well above the extrapolated value for the supercooled liquid. 
However,  we find that if we compute the difference of the chemical potential of the components with the the non-equilibrium free energy expression proposed by Jarzynski, we obtain a good match  with the extrapolated value of the supercooled liquid. The extension of the Jarzynski method that we propose opens a potentially powerful route to compute free-energy related equilibrium properties of glasses.

We find  that the  solubility estimate of amorphous materials obtained from direct coexistence simulations is only in fair agreement with the solubility prediction based on the chemical potential calculations of a hypothetical ``well-equilibrated glass''.
In direct coexistence simulations, we find that, in qualitative agreement with experiments,  the amorphous solubility decreases with time and attains a low solubility value.}
\end{abstract}
\maketitle
\section{Introduction}
Solubility is  an important   physical property of any solid. 
Solubility studies of solids are of considerable importance in environmental sciences, geochemistry (solubility of rocks), water contamination in soils, oil extraction and in the pharmaceutical industry. 
Drugs prepared in a crystalline phase often are poorly soluble in water because the stable crystal has a low chemical potential.   
A popular strategy employed in the pharmaceutical industry is  to prepare a drug in a (less stable) amorphous phase, thereby enhancing its solubility ~\cite{babu2011solubility,hancock2000true,meiron2011solubility}. 
However, the rational design of amorphous   drugs formulation  has very limited theoretical underpinning.

Glasses are formed by a rapid cooling of a liquid below its freezing temperature, which results in a dramatic increase in its viscosity whilst  avoiding crystallization \cite{angell1995formation,debenedetti2001supercooled}.  
It is  well known that glasses exhibit heterogeneous dynamics and their properties depend on the preparation protocol \cite{debenedetti1996metastable,sastry1998signatures}. 
These features of glassy dynamics affect their dissolution and long-term stability \cite{zhou2002physical,douglass2018kinetics,phan2019theoretical}, which are crucial for amorphous drug formulations.  

Parks {\it et al.} \cite{parks1934studies} reported  that for a well-annealed glucose glass, the solubility estimation from thermodynamics, which is  obtained by measuring enthalpy and entropy using the calorimetric method, matches well with the direct solubility experiments. 
The solubility of glucose glass is  around $20$ times greater than the glucose in the crystal phase. 
\textcolor{black}{The observations by Parks {\it et al.} suggest that it might be possible to compute the solubility of glasses by treating them as if they were a (metastable) equilibrium phase.}
However, for most amorphous drugs there is  no agreement of solubility values obtained from the thermodynamic data and experiments. 
In Table \ref{tab1}, we show the solubility ratios estimated from the thermodynamic data and direct experiments of glucose and other drug compounds. 
A key observation is  that from thermodynamics the solubility of the amorphous phase is  predicted to be $10$ to $1600$ times larger than  that of  the most stable crystal form.
But in experiments  the solubility of amorphous materials is  found to be much lower than what is  predicted on the basis of thermodynamics \cite{hancock2000true}. 
This discrepancy is  not very surprising, as equilibrium thermodynamics may well fail to describe the properties of (non-equilibrium) amorphous materials. 
The lower than expected solubility of amorphous materials  is  often attributed to the tendency of such materials to undergo interfacial crystallization, which would result in a decrease in the solubility~\cite{hancock2000true,janssens2009physical}. 
One common way to avoid the formation of crystals in the saturated solution is  by embedding the drug inside a polymer-based glass solution (co-amorphous drug formulation). 
Polymeric carriers stabilize the amorphous drug and improve its solubility and dissolution rate \cite{dengale2016recent}. 

\begin{table}[h] 
\caption{Solubility ratios (amorphous/crystal) for drug compounds obtained using the thermodynamic data (T) and direct experiments (E) \cite{hancock2000true}. \label{tab1}}

\begin{tabular}{|l|l|l|}
\hline
 Compound & Solubility ratio (T) & Solubility ratio (E) \\ [1ex]
 \hline
 \hline
 Glucose & $16-53$ & $24$   \\ [1ex]
 \hline
 Indomethacin & $25-104$  & $4.5$ \\ [1ex]
 \hline
 Glibenclamide & $112-1652$ & $14$  \\ [1ex]
 \hline
 Griseofulvin & $38-441$ & $1.4$  \\ [1ex]
 \hline
 Hydrochlorthiazide & $21-113$ & $1.1$  \\ [1ex]
 \hline
 Polythiazide & $48-455$ & $9.8$  \\ [1ex]
\hline
\end{tabular}
\end{table}

At present, there is, to our knowledge, no molecular-level based approach  for amorphous drug formulation.  
This problem due to the fact that amorphous solids are \textcolor{black}{kinetically arrested states and exhibit aging.}. 
Hence the tools of equilibrium statistical mechanics cannot be used to estimate their solubilities by equating the chemical potential of the molecules in the solid and in solution. 

In equilibrium,   techniques exist to compute such chemical potentials~ \textcolor{black}{\cite{bowles1996vapour}}\cite{paluch2010method,li2017computational}, but not for non-equilibrium structures such as glasses.

Computationally, the most obvious approach would be direct coexistence calculations. 
However, such an approach may be very time consuming, and less suited for materials with a low solubility.

In the present paper we explore a different approach where we use a recently developed method (the ``basin-volume approach'') to estimate the free energy that a glass would have if it could be well-equilibrated~\cite{vinutha2020numerical}.   
\textcolor{black}{This approach, although exact in principle, and accounting correctly for the configurational entropy of the glass, may fail  at very low temperatures due to inadequate sampling of the low-energy basins that dominate the properties of the equilibrium glass at these temperatures.} 
However, for the practically important case of glasses that are not cooled far below the glass transition, the approach of ref~\cite{vinutha2020numerical} should work well.

Ideally, we would like to test our approach on a glass that can be formed by slow cooling, rather than only by quenching.
However,there is  a scarcity of one-component systems of spherically symmetric particles that can form glasses upon slow cooling. 

Therefore, we test our approach on a binary glass (Kob-Andersen (KA) model) that can be made to form a glass upon slow cooling~\cite{kob1995testing,sastry2001relationship,sengupta2011dependence}. 

We combine a gradual particle insertion method with the semi-grand approach to compute the chemical potential difference between different components, and hence the solubilities of a Kob-Andersen model glass former, if the use of equilibrium arguments would be allowed.


\section{Model systems and Simulation Details}
In our study of the  binary Lennard-Jones mixture or Kob-Andersen (KA) model, we simulated $N=256$ bi-disperse spheres, 80-20 (A-B) mixture, \textcolor{black}{$N_A=204$, $N_B=52$}, density $\rho=N/V=1.2$, interacting  via 
$ V(r) = 4\epsilon_{\alpha\beta} \left[ \left(\frac{\sigma_{\alpha\beta}}{r}\right)^{12} - \left(\frac{\sigma_{\alpha\beta}}{r}\right)^6 \right] + 4\epsilon_{\alpha\beta} \left[c_0 + c_2 \left(\frac{r}{\sigma_{\alpha\beta}}\right)^2 \right]$ for $r_{\alpha\beta} < r_c$, and zero otherwise.
 Where $\sigma_{AA} = 1.0$, $\sigma_{AB} = 0.8$, $\sigma_{BB} = 0.88$, $r_c = 2.5\sigma_{\alpha\beta}$, $\epsilon_{AA} = 1.0$, $\epsilon_{AB} = 1.5$, $\epsilon_{BB} = 0.5$ 
 and $r$ is  the distance between the two pairs within in the cutoff distance \cite{sengupta2011dependence}. $c_0 = 0.01626656, c_2 = -0.001949974$ are correction terms to make the potential and force to go continuously to zero at cutoff. 
 We use simulation units: unit of length is  $\sigma_{AA}$, energy is  $\epsilon_{AA}$, mass is  $m_{A}=m_{B}=1$ and temperature is  \textcolor{black}{$\frac{\epsilon_{AA}}{k_B}$. $k_B = 1$ and $\beta = \frac{1}{k_B T}$.} 
All the thermodynamic quantities are reported in reduced units. 
Below, we report the excess free energy of the system, as the ideal gas part can be computed analytically.\\
We performed NVT Monte Carlo (MC) simulations to obtain well-equilibrated  configurations at different temperatures where the system is  not yet structurally arrested. 
To obtain glassy configurations, we perform an instantaneous quench from equilibrated liquid configurations at  $T_{H}=0.6$ for the Kob-Andersen model using the conjugate gradient minimization method \cite{press2007numerical}. We generated more than $500$ inherent structures for $T_H = 0.6$. For the direct-coexistence simulations we used $51$ glassy structures.\textcolor{black}{We use periodic boundary conditions in all our simulations}.\\

\section{Estimating Solubility}
The solubility of a solid is  the concentration of solute particles in the solvent, at coexistence.
There are two standard methods to compute solubility, \textcolor{black}{(i) direct measurement of the number of solute particles in the solvent and (ii) chemical potential based estimation (CP).}  
For solids with low solubility, the solution can be treated as an ideal solution. 
For the ideal solution, the chemical potential of the solution is  written as
\begin{equation}\label{Eq:cp_soln}
\mu_{\tiny{\text{Soln}}} = \mu_{\tiny{\text{Sov}}}^o + k_BT \ln x_{\text{s}} 
\end{equation}
\textcolor{black}{Where $\mu_{\tiny{\text{Sov}}}^o$ is the excess chemical potential of inserting one solute particle in a pure solvent}, $k_B$ is  Boltzmann's constant  and $x_{\text{s}}$ is  the mole fraction of the solute. 
In equilibrium, the chemical potential of the solute in the solid ($\mu_{\tiny{\text{Sol}}}$) phase is  equal to the chemical potential in the solution phase $\mu_{\tiny{\text{Soln}}}$. 
The solubility values from the CP method can be estimated using 
\begin{equation}\label{Eq:cp_sol}
\mu_{\tiny{\text{Sol}}}=\mu_{\tiny{\text{Sov}}}^o + k_BT \ln x_{\text{s}}.  
\end{equation}
 We use these methods to estimate the solubilities of the quenched glasses.

\subsection{Coexistence simulations}
We estimated the solubility of glasses directly by performing coexistence simulations. 
From the chemical potential method, \textcolor{black}{discussed below}, we already had estimates of the solubility, which provides a good initial guess for setting up the coexistence simulations. \textcolor{black}{We use inherent structures of the KA model as the solute. The standard Lennard-Jones model (LJ) for the solvent, with $\epsilon_{\text{S}}=1$, $\sigma_{\text{S}}=1$, and the potential is truncated and shifted to zero at $r_c=2.5\sigma_{\text{S}}$ \cite{frenkel2001understanding}. The solvent density $\rho_{\text{Sov}}=0.6$.}
We performed NVT-MC simulations for the combined system of Lennard-Jones solvent and Kob-Andersen solute. \textcolor{black}{Simulations were performed at $T=2.0$ and the solvent was kept in the fluid state.}
\textcolor{black}{We model the interaction between the solute and solvent particles with the purely repulsive Weeks-Chandler-Andersen (WCA) potential \cite{weeks1971role} and tune the interaction parameters, so that the solute particles tend to be poorly soluble in equilibrium. 
To keep the solute in the glassy state, we increase the attraction strength between the solute particles by scaling its strength with a parameter $\ell$, {\it i.e.,} $V_{\text{Sol}}= \ell V(r)$. Therefore the reduced temperature of the solute system is equal to $k_B T/\ell \epsilon_{AA}$. 
Furthermore, in order to avoid the absorption of solvent in the solute, we increase the range of the repulsive interaction between the solvent and the glass-forming particles.
$\sigma_{\text{SA}}=1.05$ for the A-type solute particles and $\sigma_{\text{SB}}=0.9$ for the B-type solute particles and $\epsilon=1$ for both types of solute particles. For the above parameters, we observe only the A component of the glass dissolves in the solvent.} 

We used a rectangular box to minimize the contact surface area between the solute and solvent. To attain equilibrium faster, we implemented swap moves in the NVT MC simulations \cite{ninarello2017models, berthier2019efficient}. 
During swap moves, we interchanged the position of a solute particle with a solvent particle. The swap moves were accepted according to the Boltzmann criterion. 
We found that successful swap moves usually involved solute particles at the interface. The swapped solute particle may not remain in the solvent for a  long time, as the successive swap moves may result in the dissolved solute particle being inserted back in the solute slab. 
Very occasionally, a solute particle would diffuse away from the interface and enter into the bulk of the solvent. 
To improve that statistics on the number of dissolved particles, we added  a bias potential, $E_{fd} = f(z)$ that increased the equilibrium concentration of solute particles in the bulk. 
The bias potential varies only in the $z$- direction. 
The potential is  zero in, and near, the solute slab, and has a finite negative value in the bulk of the solvent, $E_{fd}=-1$ (see Fig. \ref{ka_coext}a). 
Only solute particles will experience the bias:  the field does not change the solvent properties. 
By adding the bias potential we increase the probability of finding solute particles in solution.  
We correct for the effect of the bias potential and thereby obtain solubilities in the absence of the bias.
We performed swap moves $20\%$ of the time, {\it i.e.,} for every $10$ MC cycles, $2N_T$ swap moves were attempted. Each MC cycle involves $N_T=N+N_{\text{Sov}}$ trial displacement moves. We obtained the solubility estimates from run lengths of $10^6$ MC cycles, more details are presented in Appendix A.

\subsection{Basin volume method}
In what follows, we will use a method (the basin-volume approach) for estimating the effective chemical potential of well-equilibrated glasses. 
In the basin-volume approach, we can compute the free energy of glasses provided that we can reach the relevant glassy structures by energy minimization from a higher temperature where the system is  not yet glassy.
The basin-volume method computes the free energy of an equilibrium glass, even though that term seems an oxymoron. 
In principle, the basin-volume method accounts rigorously for the configurational entropy of the glass.

 From thermodynamics, we know that in equilibrium $G=\mu N$. According to the potential energy landscape picture of supercooled liquids, we can uniquely  decompose the configuration space into basins of attraction associated with different energy minima, even though we do not assume that a system in a glassy state is  necessarily trapped in such a basin (in general, it is  not).
 The configurational partition function can then be expressed in terms of a weighted sum over basin partition functions.
 Using the method of ref.~\cite{vinutha2020numerical}, we then compute the configurational partition function of such a collection of basins. 
 The advantage of the basin-volume method is  that we can compute $G$ at low temperatures, by sampling, rather than exhaustively enumerating, all basins.

Starting from a liquid state at a high-temperature $T_H$, we perform a fast quench using the conjugate gradient energy minimization, to find energy minima  (``inherent structures'' (IS)). 
The initial configurations of the quench are obtained by Monte Carlo sampling of the liquid and are therefore Boltzmann weighted. 
To obtain configurational free energy for low-temperature ($T_L$) glasses, we need to perform a large number of thermodynamic integrations (TI) where we cool the system confined to a given basin  from $T_H$ to $T_L$. 

{\color{black}The Basin Volume method is based on the observation that the equilibrium partition function of a glass can be written rigorously as:
\begin{equation}\label{eq:QL_from_QH}
Q(T_L)=\sum_i q_B^i(T_L) = Q(T_H)\times \sum_i \left({q_B^i(T_H)\over Q(T_H)}\right)\left( {q_B^i(T_L)\over q_B^i(T_H)}\right)
\end{equation}
where the sum runs over {\em all } basins, and $q_B^i(T)$ denotes the partition function of the system confined to a given basin $i$ at temperature $T$.
As all basins together span the entire configuration space, the first equality in Eqn.~\ref{eq:QL_from_QH} is obvious. 
The subsequent ones are obtained trivially by multiplying and dividing by the same factors.

There is no need to evaluate   $q_B^i(T_H)/Q(T_H)$  because this ratio is simply equal to  the probability $P_i$ that basin $i$ is sampled (at $T_H$).

When we perform an MC simulation of the liquid at $T_H$,  we visit the $i^{th}$ basin  with a probability $q_B^i(T_H)/Q(T_H)$.

Then:
\begin{equation}\label{eq:QL_from_QH2}
Q(T_L)= Q(T_H)\times \left\langle {q_B^i(T_L)\over q_B^i(T_H)}\right\rangle_{MC}
\end{equation}
or
\begin{equation}\label{eq:FL-FH}
\beta_L F_L = \beta_H F_H  -\ln \left\langle {q_B^i(T_L)\over q_B^i(T_H)}\right\rangle_{MC}
\end{equation}

Note that the effect (and advantage) of this algorithm is that it  replaces an {\em enumeration} over basins $i$ by sampling. 

The key quantity to compute (by thermodynamic integration) is the ratio $q_B^i(T_L)/q_B^i(T_H)$, for a sample of states $i$. 
Of course, we also need to know $F_H$ for the equilibrium liquid at $T_H$, but that only  involves standard thermodynamic integration~\cite{frenkel2001understanding}.
For details, see ref.~\onlinecite{vinutha2020numerical}.}

For a system of a single component, the basin-volume method provides an estimate of the solubility chemical potential of the equilibrated glass, and thereby of its solubility, provided that we know the excess chemical potential of the solute in solution. 
The latter quantity can be computed using standard methods (see e.g.~\cite{frenkel2001understanding,li2017computational}).
For a multi-component system, where $G = \mu_1 N_1 + \mu_2 N_2 +..$, the basin-volume method is not enough to determine the solubility of a given species.  For a 2-component system,  we also need to know the difference in chemical potentials $\Delta \mu = \mu_1 - \mu_2$ of the two species.
In the next subsection, we describe a method to compute the difference in chemical potential at low temperatures. 
 
\subsection{\textcolor{black}{Thermodynamic integration of interaction parameters}}
 The Gibbs free energy per mole is given by
\begin{eqnarray}
G(\chi_{A}) = \chi_{A}\mu_A + \chi_{B}\mu_B
\end{eqnarray}
Where $\chi_A (= 1-\chi_B)$ denotes the mole fraction of species A. We can compute $\Delta \mu$ , using:
\begin{eqnarray}
\left( \frac{\partial G(\chi)}{\partial \chi} \right)_{N,P,T} = \mu_A - \mu_B
\end{eqnarray} 
Even though standard computational tools such as the Widom particle-insertion method and semi-grand ensemble method \cite{frenkel2001understanding} are valid in principle, these methods will not provide reliable estimates in the case of dense liquids. 
We therefore use a method, which is
a combination of the semi-grand ensemble method and a gradual particle insertion method \cite{mon1985chemical} and then perform thermodynamic integration to compute $\Delta \mu$. 
This method can be used at any density and temperature for a system in equilibrium. 
The basic idea is  simple, we transform a B-type particle to an A-type particle by slowly (and hopefully reversibly) changing the diameter $\sigma$ and $\epsilon$, by introducing a $\lambda$ parameter we perform thermodynamic integration to compute the change in the free energy or $\Delta \mu$.
We pick a particle $i$ of B type and then transform to A-type particle, as shown below:
\begin{eqnarray}
 \sigma_{iA}^{6} &=& (1-\lambda)\sigma_{AA}^{6} + \lambda \sigma_{BA}^{6}  \\\label{eqslins6}
 \sigma_{iB}^{6} &=& (1-\lambda)\sigma_{AB}^{6} + \lambda \sigma_{BB}^{6}   \\
 \epsilon_{iA} &=& (1-\lambda)\epsilon_{AA} + \lambda \epsilon_{BA}  \\
 \epsilon_{iB} &=& (1-\lambda)\epsilon_{AB} + \lambda \epsilon_{BB}   
\end{eqnarray}
From the above equations, $\lambda=1$ corresponds to the system of $N_A, N_B$ particles and $\lambda=0$ corresponds to the system of $N_A+1,N_B-1$ particles. Therefore, the change in free energy is  given by
\begin{eqnarray}\label{eq:delfka}
 F(\lambda=1) - F (\lambda=0) =   \int_{\lambda=0}^{\lambda=1} \left\langle \frac{\partial u_i(\lambda)}{\partial\lambda}\right\rangle_{\lambda} d\lambda   
\end{eqnarray}
\begin{equation}
\begin{split}
 \frac{\partial u_i(\lambda)}{\partial\lambda} =& \frac{4\epsilon(\lambda)}{\sigma^6(\lambda)}\left[2\left(\frac{\sigma(\lambda)}{r}\right)^{12}-\left(\frac{\sigma(\lambda)}{r}\right)^{6}\right] \left(\frac{\partial \sigma^6(\lambda)}{\partial \lambda}
\right) \\
& + 4 \left[\left(\frac{\sigma(\lambda)}{r}\right)^{12}-\left(\frac{\sigma(\lambda)}{r}\right)^{6}\right] \left(\frac{\partial \epsilon(\lambda)}{\partial \lambda}\right)
\end{split}
\end{equation}\label{equlam}
\textcolor{black}{where $\langle ...\rangle_{\lambda}$ denotes an ensemble average of the $i^{th}$ particle potential energy $u_i(\lambda)$}. We performed NVT MC simulations for $10$ values of $\lambda$ and then performed a $10-$point Gauss quadrature to compute the free-energy difference. 
To obtain accurate estimates for the Kob-Andersen model, we varied $\sigma^6$ (rather than $\sigma$) (Eq.\ref{eqslins6}) linearly with $\lambda$, so that the integrand in the Eq.\ref{eq:delfka} changed slowly with $\lambda$. \textcolor{black}{From Eq. \ref{eq:delfka} and the thermodynamic relation $\Delta G = \Delta F + \Delta (PV)$ we obtain the change in the Gibbs free energy using the fact that the dense phase is barely compressible. Then we can replace $\Delta (PV)$ by $V \Delta P$. We integrate the pressure difference  $\Delta P = P(\lambda=1)-P(\lambda=0)$.
Then $\Delta G$= $\Delta F$ + $V\Delta P$, where $V = 1/\rho$ is the volume per particle.} 
We validated our method by computing the chemical potential for the binary Lennard-Jones model at the same state point as that in the paper of Perego {\it et al.}  \cite{perego2016chemical}. We find $\mu^{A} = 4.8$ which is  consistent with the value $\mu^{A} = 4.1$ reported in Perego {\it et al.} paper \cite{perego2016chemical}. 
Our value is  averaged over $10$ independent runs, with a standard deviation of $0.15$. We found that the above method is  simpler to implement and more accurate than the corresponding chemical potential estimates obtained using metadynamics \cite{perego2016chemical}.  

In the results section, we show the chemical potential estimates for the Kob-Andersen model at high $T$ and in a supercooled state. 
Also, using the above method and the basin-volume method, we attempt to obtain estimates of $\mu_A$ and $\mu_B$ and their solubilities for the Kob-Andersen glasses. \\

We use the above method to also compute $\mu_{\tiny{\text{Sov}}}^o$. \textcolor{black}{We perform NPT MC simulations of the solvent, at $\rho_{Sov}=0.6, T=2, P=2$.} In our study, solvent particles interact via the Lennard-Jones potential and the interaction between solute and solvent particles is  the WCA potential. 
We can split the Lennard-Jones potential ($U^{\text{LJ}}(r)$) into a repulsive part $U^{\text{WCA}}(r)$, and an attractive part $W(r)$.
\begin{equation}
U^{\text{LJ}}(r) = U^{\text{WCA}}(r) + W(r)
\end{equation}
We can further write it as
\begin{eqnarray}
U^{\text{WCA}}(r) &=& U^{\text{LJ}}(r) + \epsilon, r < 2^{1/6}\sigma \\
	&=& 0, r \geq 2^{1/6}\sigma \\
W(r) &=& -\epsilon,  r < 2^{1/6}\sigma \\
	&=& U^{\text{LJ}}(r), r \geq 2^{1/6}\sigma 
\end{eqnarray}
We pick a solvent particle $i$ and convert it to a solute particle.
 We change the diameter of a solvent particle ($\sigma_{\text{Sov}} = \sigma_{AA}$) to a solute particle of $\sigma_{\textcolor{black}{SA}}$, using the following equation: 
\begin{equation}
 \sigma_{i}^{6} = (1-\lambda)\sigma_{\textcolor{black}{SA}}^{6} + \lambda \sigma_{\text{Sov}}^{6}  \label{eq:sovsol}
\end{equation}
For the interaction potential, we reversibly switch off the attractive part of the interaction potential to change from the solvent-solvent interaction to the solvent-solute interaction for the $i^{\text{th}}$ particle, as follows:
\begin{equation}
 u_{i}(r) = u_i^{\text{\tiny{WCA}}}(r) + \lambda w_i(r)
\end{equation}
\begin{equation}
\begin{split}
\frac{\partial u_i}{\partial\lambda} =& (1+\lambda)\frac{4\epsilon}{\sigma_i^6(\lambda)}\left[2\left(\frac{\sigma_i(\lambda)}{r}\right)^{12}-\left(\frac{\sigma_i(\lambda)}{r}\right)^{6}\right] \left(\frac{\partial \sigma_i^6(\lambda)}{\partial \lambda}
\right) \\
&+ w_i(r)  \label{eq:usolsov}
\end{split}
\end{equation}
From the above equations, $\lambda = 1$ corresponds to the pure solvent system and $\lambda =0$ corresponds to the solvent system with one solute particle. 
We performed NPT MC simulations for $10$ values of $\lambda$  and computed $\left\langle \frac{\partial u_i}{\partial\lambda} \right\rangle_{\lambda}$. 
Using Eq. \ref{eq:delfka}, we obtained the change in the Gibbs free energy is equal to $\mu_{\text{Sov}}^o - \mu_{\text{Sov}}^{\text{LJ}}$. \textcolor{black}{Using the known chemical potential value of the Lennard-Jones system $\mu_{\text{Sov}}^{\text{LJ}} \approx 0.3$ \cite{frenkel2001understanding}, at the state point $\rho_{Sov}=0.6, T=2, P=2$, we estimated the value of $\mu_{\text{Sov}}^o = 8.96$ for the A-type solute particle.}

\section{KA glasses}
\begin{figure}[h!]
\includegraphics[scale=0.35]{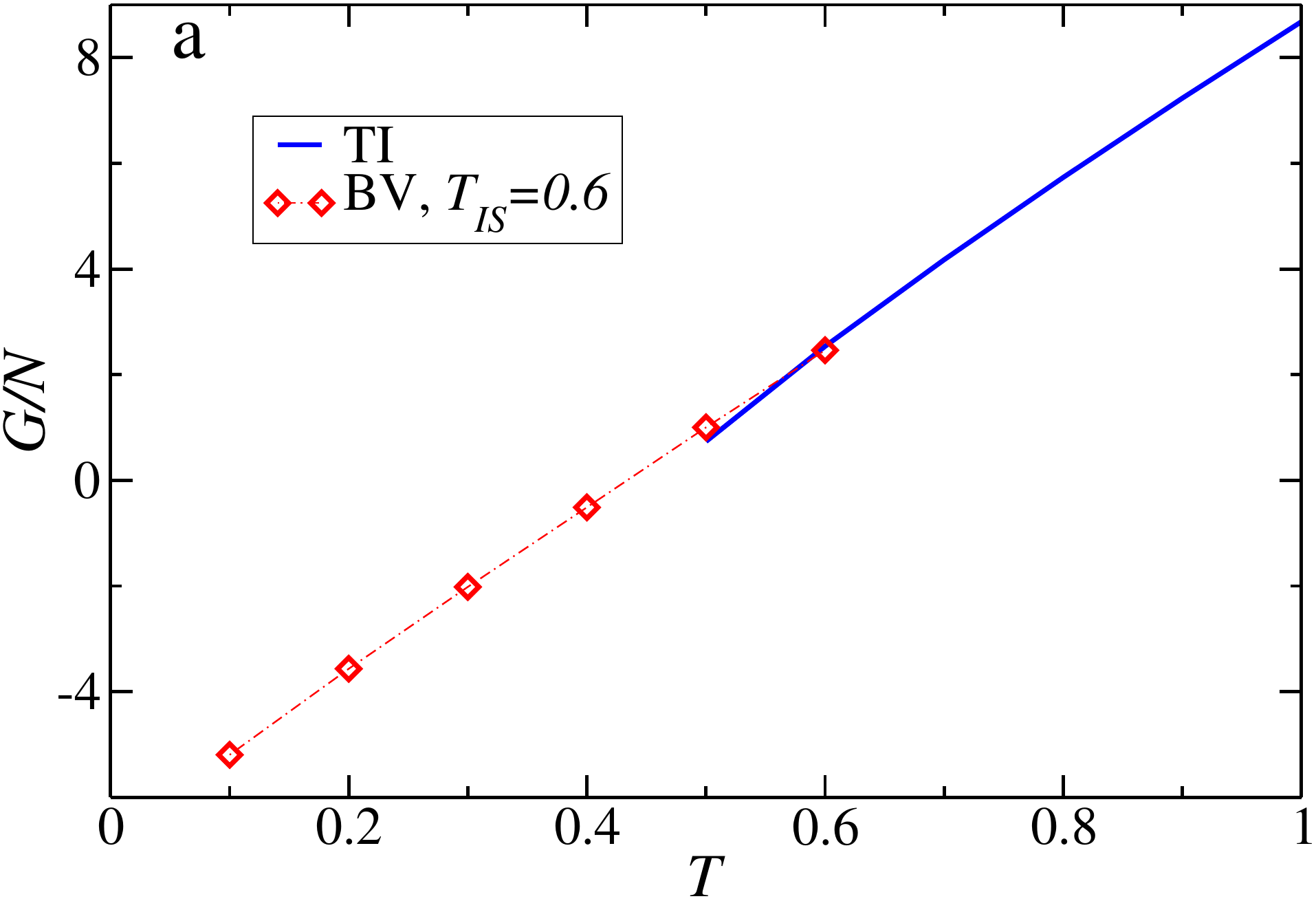}
\includegraphics[scale=0.35]{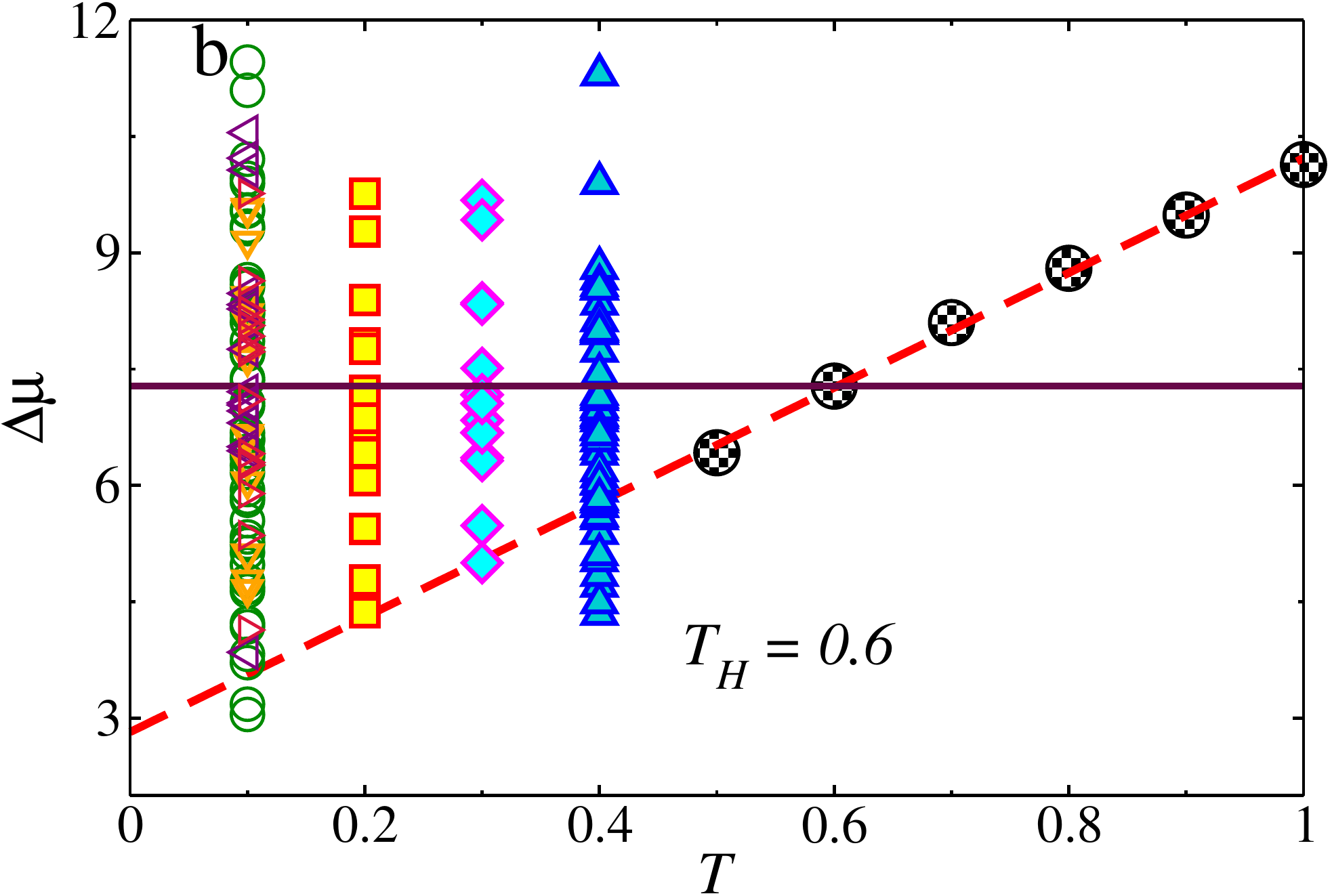}
\includegraphics[scale=0.35]{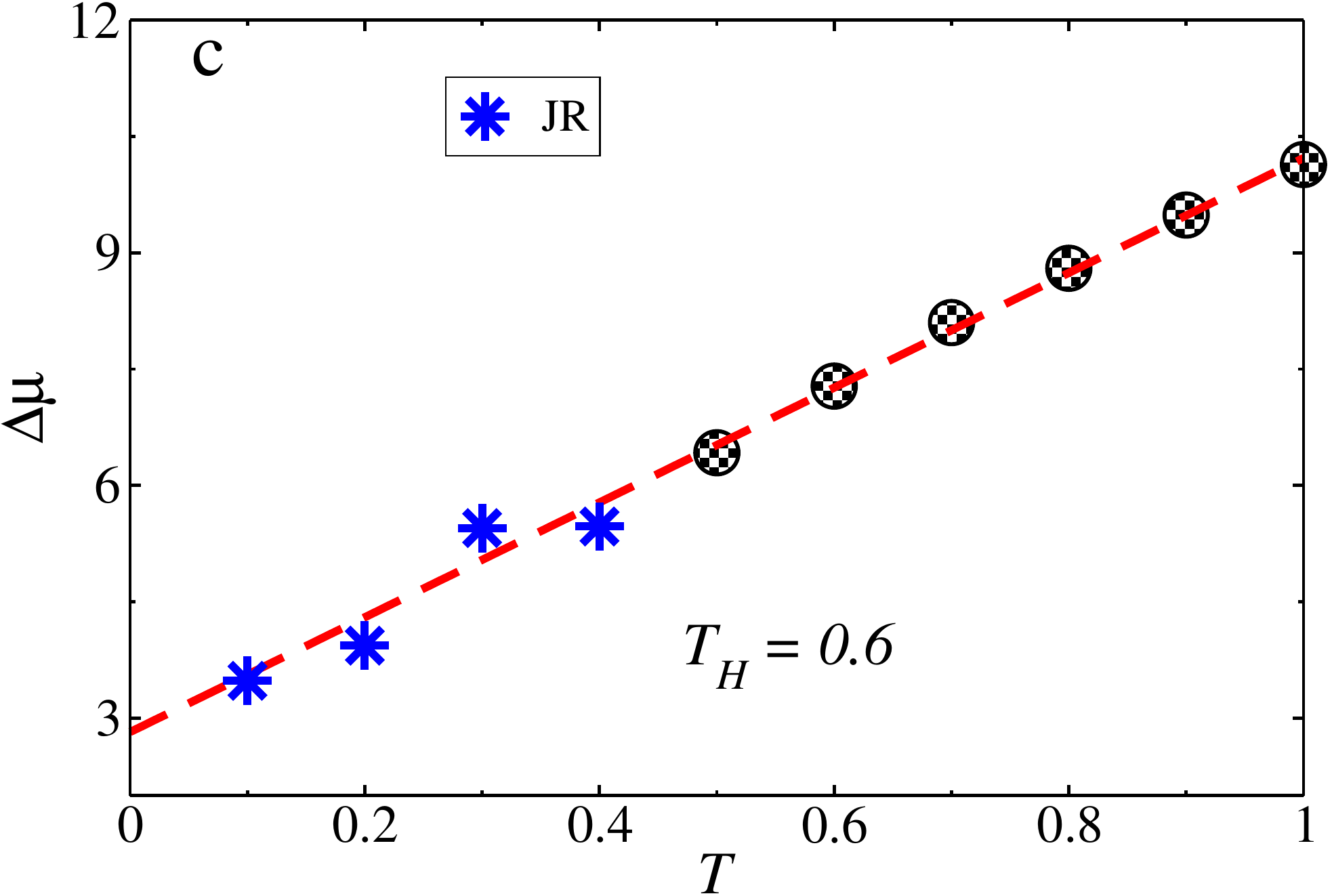}
\caption{\label{KA_cp}{\bf (a)} The Gibbs free energy per particle shown as a function of temperature, for the glasses at $T_{\text{H}}=0.6$. {\bf(b)} $\Delta \mu$ as a function of temperature, which is  computed using  thermodynamic integration. 
For the glass configuration at $T=0.2-0.4$, each data point corresponds to different B-type particles transformed to A-type particles in the same initial configuration. 
For $T=0.1$, we show data for four glass configurations. The horizontal bold line corresponds to the average $\Delta \mu$ for one inherent structure (shown as circles), averaged over different $B$ particles in the same initial configuration. 
The dashed line is a linear fit to the supercooled liquid data. \textcolor{black}{{\bf (c)} $\Delta \mu$ for the glass configuration at $T=0.1-0.4$, which is computed using  Jarzynski's relation (JR).}}
\end{figure}

In this section, we present solubility estimates for quenched, amorphous solids. 
To obtain glassy inherent structures,  we quenched configurations of a  (supercooled) liquid well-equilibrated at temperature $T_{\text{H}} = 0.6$. 
In Fig.~\ref{KA_cp}(a), we show $G$ for the low-$T$ glasses, obtained using the basin-volume method, and liquid configurations obtained using TI. 
To estimate $\mu_{\text{A}}$ and $\mu_{\text{B}}$ of the binary Kob-Andersen mixture, we compute $\Delta \mu = \mu_{\text{A}} - \mu_{\text{B}}$ using  thermodynamic integration, as discussed in section IIIB. In Fig.~\ref{KA_cp}(b), we show $\Delta \mu$ for the equilibrium (supercooled) liquid at temperatures $T=1.0-0.5$ and the low-$T$ configurations with $T_{\text{H}}=0.6$. 
For the low-$T$ glasses with $T_{\text{H}}=0.6$, we performed NVT MC starting with initial configurations obtained using the basin-volume method and compute $\Delta F$ using Eq.~\ref{eq:delfka}. 
Each data point in Fig.~\ref{KA_cp}(b) for the low-T configurations corresponds to the result of the thermodynamic integration for a single B-type particle, which is  transformed to a A-type particle. The data for  $T>0.1$ were obtained by selecting different $B$ particles in the same configuration
However, for $T=0.1$, we also started runs from different initial configurations. 
In Fig.~\ref{KA_cp}(b), we use different symbols to distinguish the $\Delta \mu$ for different initial  configurations. 
For every  thermodynamic integration simulation we used $10$ values of $\lambda$ and $2 \times 10^6$ MC cycles for each value of $\lambda$. 
For the liquid configurations, the number of samples, MC cycles and standard deviation are \{$(10,2 \times  10^6,0.2)$,$(20,2 \times 10^6,0.08)$,$(10,2 \times 10^6,0.05)$,$(10,10^6,0.1)$,$(10,10^6,0.1)$, $(10,10^6,0.07)$\} for temperatures \{$0.5,0.6,0.7,0.8,0.9,1.0$\}, respectively.  
\textcolor{black}{However,  for the glassy configurations at $T=0.1,0.4$, the values of $\Delta \mu$ is  quite broad. } 
The fact that we obtain a distribution of $\Delta \mu$s for $T\le 0.4$ is  an indication that the system is  no longer in equilibrium. 
This observation is  important, because it means that the concept of {\em the} chemical potential of a species is  no longer meaningful: different $B$-particles experience different environments, from which they do not escape on the time-scale of a simulation. 

We note, however,  that the average value of $\Delta \mu$ at $T=0.1$ and $T=0.4$  is  close to the value for the corresponding liquid at $T=0.6$. 
At $T=0.4$ the $B$ particles can explore some different environments and, as a consequence,  distribution of $\Delta \mu$ is  somewhat narrower for longer runs. No such narrowing is  observed  in the case of $T=0.1$. 

To check for obvious structural properties that might correlate with $\Delta \mu$, we computed the local bond-orientational order parameter~\cite{steinhardt1983bond} and the initial energy of the different B-type particles. However, we did not observe any strong correlation of either quantity with $\Delta \mu$. 
  
The fact that there is  a wide distribution of  $\Delta \mu$ values for low-$T$ glassy structures  (Fig. \ref{KA_cp}(b))  implies that different particles of the same type would have different propensities to dissolve. 
This means that the very concept of {\em the} solubility of glassy materials is  questionable.
As such, this finding may not be surprising owing to the non-equilibrium nature of glasses, but we are not aware of earlier numerical data  that illustrate this point. 
As particles in different environments have different excess chemical potentials, one would expect the solubility of a glass to depend on  the distribution of solute chemical potentials in the glass and on  the distribution of chemical potentials at the interface (i.e. of those particles that are kinetically accessible from the liquid).
This distribution may be different from that in the bulk of the glass.
In particular the latter factor may be relevant for the time-dependent solubility properties  of amorphous drugs, see Table. \ref{tab1}. 

\begin{figure}[h!]
\includegraphics[scale=0.35]{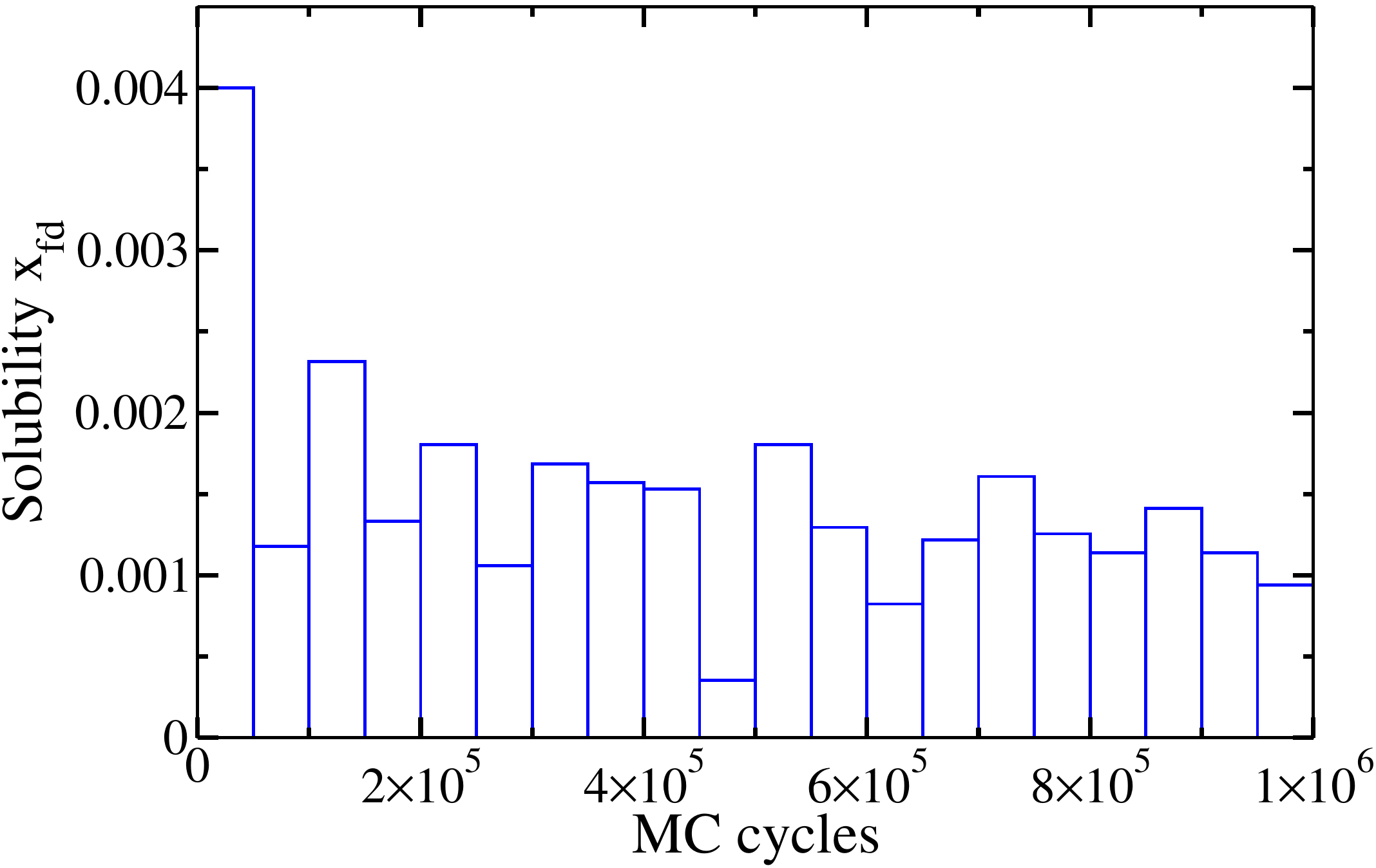}
\caption{\label{KAsolu_time} Amorphous solubility as a function of MC cycles, which is  computed in the presence of the field $E_{fd}=-1$. Number of samples used for averaging is  $51$.}
\end{figure}

Having indicated the problems of computing chemical potentials in glasses, we nevertheless explore to what extent we can estimate  the solubility of glassy materials using the CP method by  comparing these predictions with the solubility estimates obtained from  direct coexistence simulations (DC).\\
\textcolor{black}{For the low-T glasses, the distribution of $\Delta \mu$ is wide. Therefore, to obtain the average $\Delta \mu$, we use the non-equilibrium free energy expression due to Jarzynski \cite{jarzynski1997nonequilibrium,jarzynski1997equilibrium}. Jarzynski's relation (JR) is as follows:
\begin{equation}\label{eq:JR}
\exp(-\beta \Delta F) = \overline{\exp[-\beta W(t_s)]}
\end{equation}
The above equation relates the free energy difference between two systems to the non-equilibrium work ($W$) needed to transform one system into the other in an arbitrarily short ``switching" time ($t_s$). It implies that we can obtain information about equilibrium free energy differences from a non-equilibrium simulation. In the limit of infinitely slow switching, the system remains in equilibrium, we recover the relation between $\Delta F$ and the reversible work $W_s$, $\exp(-\beta \Delta F) = \exp(-\beta W_s)$. Here, for $T=0.1-0.4$, we use Eq. \ref{eq:delfka} to obtain the non-equilibrium work needed to transform a B-type particle to an A-type particle. Using Eq.\ref{eq:JR}, we obtain the average $\Delta\mu$ for the low-$T$ glasses, see Fig.\ref{KA_cp}(c). The number of samples are \{$97,46,46,52$\} for temperatures \{$0.1,0.2,0.3,0.4$\}, respectively. Surprisingly, we observe that $\Delta\mu$ obtained using Jarzynski's relation matches very well with the extrapolated $\Delta \mu$ value of the supercooled liquid branch (the dashed line in Fig. \ref{KA_cp}(c)). 
We obtain estimates of $\mu_{\text{A}}^g$ and $\mu_{\text{B}}^g$ for the low-$T$ glasses using $\Delta\mu = \mu_{\text{A}} - \mu_{\text{B}}$ from Jarzynski's relation and  the average chemical potential ($G = N_A \mu_{\text{A}} + N_B \mu_{\text{b}}$), from the basin volume method.
}
\begin{table}[h]
\caption{\textcolor{black}{Comparison of solubility values obtained for a single reference IS using the chemical potential (CP) method and direct coexistence (DC) simulations. We show the solubility estimates for A type solute particles from the DC simulations for different values of attraction strength $\ell$ and the CP at different temperatures.}\label{KA_solub}}
\begin{tabular}{|l|l|}
\hline
$x_s$, \textcolor{black}{$k_BT/\ell\epsilon_{AA}$} (DC) & $x_s$, \textcolor{black}{$k_BT/\epsilon_{AA}$} \textcolor{black}{(CP)} \\ [1ex]
 \hline
 $0.018$, $0.3$ & $0.007$, $0.3$   \\ [1ex]
 \hline
 $1.2 \times 10^{-3}$, $0.2$ & $2.9 \times 10^{-3}$, $0.2$  \\ [1ex]
 \hline
 $2.2 \times 10^{-6}$, $0.1$  & $1.2 \times 10^{-3}$, $0.1$  \\ [1ex]
\hline
\end{tabular}
\end{table}
Starting with the inherent structure at $T_{\text{H}}=0.6$ (the same inherent structure for which $\Delta \mu$ was estimated), we performed NVT MC for the solute/solvent system, with $N_{\text{Sov}}=500$,$N=256$,$T=2.0$ \& $\ell=10$, see Appendix A for more details.
In Table \ref{KA_solub}, we show for a reference inherent structure the solubilities obtained from the CP and direct-coexistence method. \textcolor{black}{Using $G$ (Fig. \ref{KA_cp}(a)), $\Delta \mu$ of the KA glass at a given temperature, we obtain $\mu_{\text{A}}^g$. we insert $\mu_{sol} = \mu_{\text{A}}^g$ and $\mu_{\text{Sov}}^o$ in Eq.\ref{Eq:cp_sol}, we get an estimate for $x_s$.
In the direct-coexistence (DC) simulations, we directly measure the solubility of the solute slab at different temperatures by varying the attraction strength $\ell$ between the solute particles.} 
In the relatively small samples that we use, we observe pronounced sample-to-sample variation in the solubility. 
However, we observe that the average DC solubility values are comparable to the values obtained from the extrapolation of the supercooled-liquid branch for $T=0.3,0.2$. 
However, for $T=0.1$, the chemical potential method fails to provide a reasonable estimate. 
This failure is probably due to existence of low energy inherent structures that are accessed in DC simulations, but not in the chemical potential approach. 
For this reason, the chemical potential approach should not be used for glasses at very low temperatures.
It is  interesting to note that the average chemical potential of $A$ particles in the bulk glass does {\em not} yield a good prediction of the solubility.
This finding suggests that the $A$ particles that can exchange with the liquid have (on average) a lower chemical potential than those in the bulk.
This finding is not surprising: solute  particles at the glass-liquid interface can diffuse and find regions with lower $\Delta \mu$. 
In the bulk, such annealing is not possible. 
The fact that the solubility of the particles at the surface are comparable with the value obtained by extrapolating the chemical potential of the well-equilibrated supercooled liquid, suggests that the mobile solute particles at the interface can, in fact, equilibrate.\\ 
Evidence for the hypothesis that solute particles at the interface are equilibrating comes from the fact that in experiments the solubility of amorphous materials  appears to decrease with time and before reaching a plateau value \cite{hancock2000true,douglass2018kinetics}. 
We observe the same behavior in our direct-coexistence simulations (see Fig.~\ref{KAsolu_time}).
In experiments on the dissolution of  amorphous drugs, the high initial dissolution rate is  followed by recondensation. 
It is  commonly assumed that this recondensation results in the formation of crystalline layers on the surface. 
As the crystal is  more stable than the amorphous phase, crystallization would result in a decrease of the solubility of the drug. 
Although, in our model system, we do observe that the solubility decreases with time, we do not observe any crystallization on the solute slab.
\section{Discussion}
To summarize: we have explored a numerical method to compute solubilities of amorphous solids. 
\textcolor{black}{For supercooled liquid configurations, we use a thermodynamic integration method   and for the low-T glasses, we employ Jarzynski's free energy expression to compute the chemical potential difference between the two components of the Kob-Andersen model. We show that the use of Jarzynski's relation provides a new way to compute the equilibrium properties of glasses.
We use the non-equilibrium free energy method and the basin-volume method to estimate the solubilities for the Kob-Andersen model glass former.} 
\textcolor{black}{A key finding of our study is that the chemical potential method, which works well for equilibrium phases, fails to provide solubility estimates for glasses at low temperatures.}\\
From the direct coexistence simulations, we show that the solubility of glass decreases with time and attains a low value. 
{\color{black} To our knowledge, this approach to compute the free-energy change associated with the change of the Hamiltonian in a glassy system, amounts to a new application of the Jarzynski method.
Normally, the Jarzynski method is equivalent to thermodynamic integration. However, in the present case, thermodynamic integration fails, while the Jarzynski method still works.}

The solubility values from the direct coexistence simulations are comparable to those of a well-equilibrated glass \textcolor{black}{at $T=0.3,0.2$}. 

We observe the memory of the preparation protocol in the average value of the chemical potential of the quenched glasses. 
The local chemical potential in the quenched glass results in particles having different propensity to dissolve\textcolor{black}{, suggesting that, at the interface, the particles with a large excess chemical potential will dissolve preferentially, whilst particles with a lower chemical potential will be deposited. 
This process may result in the interface layer equilibrating faster than the bulk of the glass.}
\textcolor{black}{Developing a simple single-component glass former is vital to gain insights into amorphous solubility as it becomes amenable to computational approaches like the basin volume method and coexistence simulations.}
Our approach can be extended to study the role of structural and dynamical heterogeneity of glasses in determining its dissolution, which is  essential for rational amorphous drug formulations. 
The dependence of local structure on the chemical potential is  a key factor in many applications related to etching and failure in glasses, and in designing of ultrastable glasses.  

\begin{acknowledgments}
We gratefully acknowledge the funding by the International Young Scientist Fellowship of Institute of Physics (IoP), Chinese Academy of Sciences under grant no. 2018008. We gratefully acknowledge IoP and the University of Cambridge for computational resources and support. HAV acknowledge very useful discussions with Srikanth Sastry, and Jure Dobnikar.
\end{acknowledgments}

\section*{Data availability}
The data that support the findings of this study are available from the corresponding author upon reasonable request.

\appendix
\section{Coexistence simulations of Kob-Andersen solute in Lennard-Jones solvent}
\begin{figure}[h!]
\includegraphics[scale=0.25]{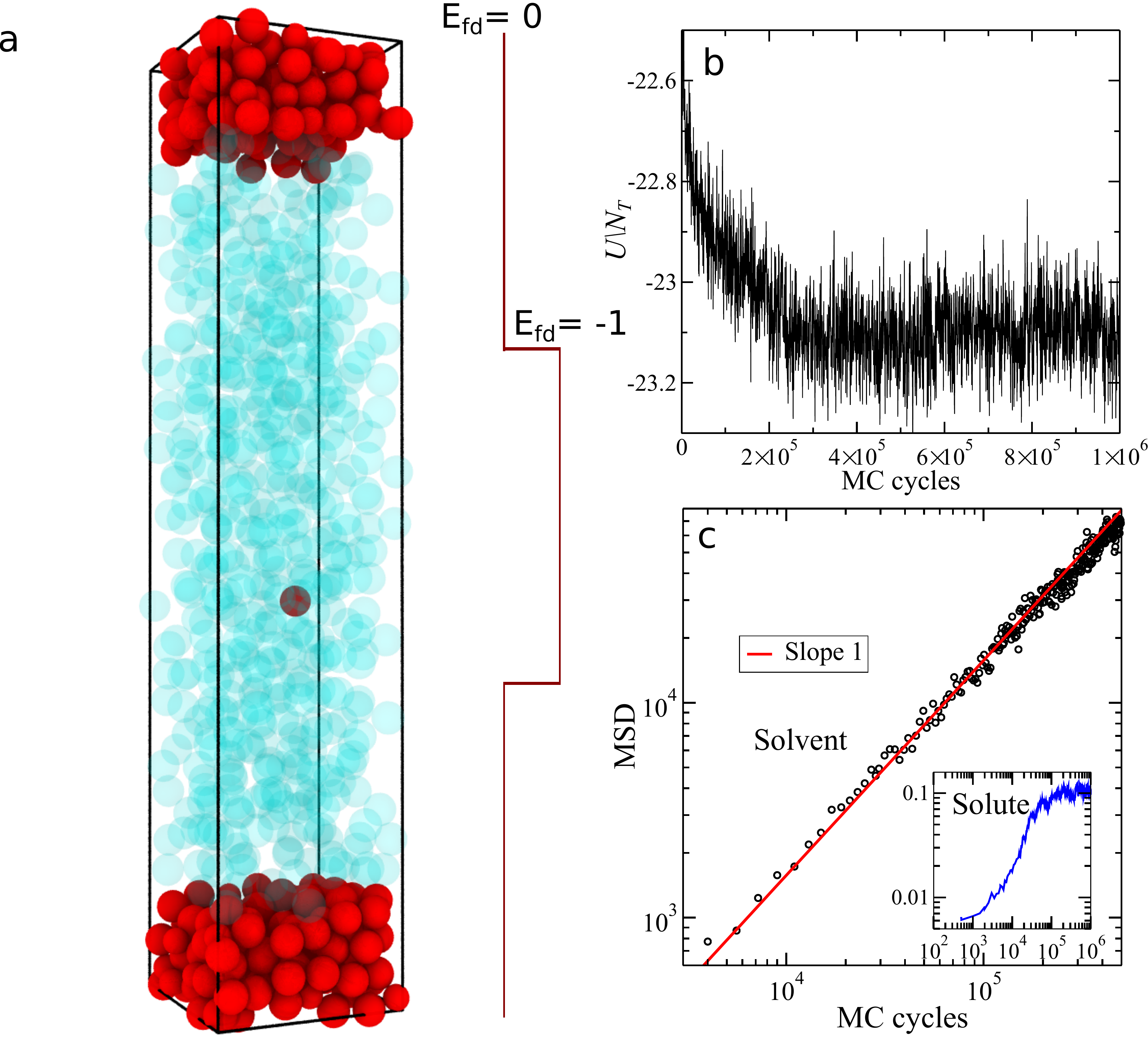}
\caption{\label{ka_coext} Coexistence simulations: {\bf (a)} Snapshot of the coexistence simulations of the Kob-Andersen solute (opaque red spheres) in the Lennard-Jones solvent (transparent blue spheres), with the external step potential $E_{fd}=-1$. 
We observe one solute particle dissolved in the solvent. {\bf (b)} The potential energy of the combined systems shown as a function of MC cycles. 
The system attains equilibrium around $3 \times 10^5$ MC cycles. {\bf (c)} MSD as a function of MC cycles for the solvent and solute particles in the bulk (inset). 
The solvent shows diffusion and the solute behaves as a glass.}
\end{figure}

We perform NVT MC swap coexistence simulations, in the presence of the external step potential, for a combined system of the Kob-Andersen solute and Lennard-Jones solvent. 
To minimize the interface between the solute and solvent, we use a rectangular simulation box. 
Box lengths along $x$- and $y$- direction is  fixed by the density of the solute, {\it i.e.,} $L_x=L_y=(\frac{N}{\rho})^{\frac{1}{3}}$. 
The box length along $z$- direction is  given by $L_z = (\frac{N_{\text{Sov}}}{\rho_{\text{Sov}}})^{\frac{1}{3}} +L_x$. 
The initial configuration for the solute is  the inherent structure at $T_{\text{H}}=0.6$ and a random initial configuration for the solvent at $\rho=0.6$ and $T=2$, we perform $10^2$ MC cycles for initial equilibration of the Lennard-Jones solvent before we attempt swap moves. 
The MC simulations are performed in the presence of the external step potential, which has a value of $-1$ in the region \{$(L_x,L_y,L_z=L_x)$,$(L_x,L_y,L_z=-L_x)$\} from the center of the simulation box and zero otherwise, see Fig. \ref{ka_coext}a. 
We tune the inter-particle interactions for the combined system such that we obtain low-solubility but finite, see section II. 
The external step potential aided in the dissolution of one or two solute particles in the bulk of the solvent and to obtain the solubility estimates. 
In Fig. \ref{ka_coext}b, we show the evolution of potential energy per particle as a function of MC cycles for the combined system. In Fig. \ref{ka_coext}c, we show the mean squared displacement (MSD) of the solvent and solute in the figure inset. 
It is  clearly seen that the MSD of the solvent is  diffusive and the solute is  glassy. 
For the solute, only the MSD of particles in the bulk of the solute is  considered as the solute particles on the surface can diffuse and dissolve. 
Dissolved solute particles are those which enter the bulk region with $E_{fd}=-1$. 
We mostly observe one dissolved solute particle at any time. 
To estimate solubilities, we count only the dissolved solute particles after the system has reached equilibrium, {\it i.e.,} after $3\times 10^5$ MC cycles. 
For the reference inherent structure for which $\Delta \mu$ (green circles) at $T=0.1$ was estimated in Fig. \ref{KA_cp}(b), the solubility value $x_{fd}= 0.002$ ($1$ solute particle). 
We use the $x_{fd}$ value to compute the required zero-field solubility, which is  given by $x_s = x_{fd}\exp(E_{fd}/RT)$. 

\bibliography{solubility_glass_v3.bib}
\end{document}